\documentstyle[aps]{revtex}
\def\be{\begin{equation}}
\def\ee{\end{equation}}
\def\ba{\begin{array}}
\def\ea{\end{array}}
\def\bea{\begin{eqnarray}}
\def\eea{\end{eqnarray}}
\def\nn{\nonumber}

\def\p{\partial}
\def\pr{\p_r}
\def\pv{\p_v} 
\def\pta{\p_{\theta}}
\def\pvi{\p_{\varphi}}
\def\cD{\cal D}
\def\cL{\cal L}
\def\cLd{{\cal L}^{\dagger}}

\def\sta{\sin\theta}
\def\cta{\cos\theta}
\def\sda{\sin^2\theta}
\def\cda{\cos^2\theta}
\def\coa{\cot\theta}
\def\sqd{\sqrt{2}}
\def\hf{1/2}
\def\prH{r_H^{\prime}}
\def\pdrH{r_H^{\prime 2}}
\def\pprH{r_H^{\prime\prime}}

\begin{document}
\pagestyle{myheadings}
\markboth{}{Extra Spin-Rotation Coupling Effect in 
a Radiating Kerr Space-time \hfill Wu and Cai~~~~}
\draft
\twocolumn[\hsize\textwidth\columnwidth\hsize\csname 
           @twocolumnfalse\endcsname

\title{\bf Extra Spin-Rotation Coupling Effect in
a Radiating Kerr Space-time}
\author{S. Q. Wu\thanks{E-mail: sqwu@iopp.ccnu.edu.cn} and 
X. Cai\thanks{E-mail: xcai@ccnu.edu.cn} }
\address{Institute of Particle Physics, Hua-Zhong 
Normal University, Wuhan 430079, China}
\date{\today}
\maketitle

\begin{abstract}
\widetext
Source-less wave equations are derived for massless scalar, 
neutrino and electromagnetic perturbations of a radiating 
Kerr space-time, and the Hawking radiation of massless
particles with spin $s =0, 1/2$ and $1$ in this geometry 
is investigated by using a method of the generalized tortoise 
coordinate transformation. An extra interaction between the 
spin of particles and the rotation of the hole displays 
in the thermal spectra of Hawking radiation of massless 
particles with spin $s = 1/2, 1$ in the evaporating Kerr 
space-time. The character of such effect is its 
obvious dependence on different helicity states of 
particles with higher spin.
\end{abstract}
\pacs{PACS numbers: 97.60.Lf, 04.70.Dy} 
]

\narrowtext

It is generally accepted in black-hole physics that the total 
mass-energy of a rotating charged black hole can be separated 
into three parts \cite{CR}: the rotational energy, the Coulomb 
energy and the irreducible mass, which was reinterpreted later 
by Smarr \cite{Smar} as the surface energy. Correspondingly, the 
energy of a charged particle in a charged axisymmetric black hole 
is composed of the electromagnetic interaction part and the energy 
due to the coupling of the orbital angular momentum of a particle 
with the rotation of the black hole. Besides these two ingredients, 
one can expect naively that there are other interaction 
energies such as those arising from the gravitational coupling 
of rotation or acceleration of a black hole with the intrinsic spin 
of a spinning particle \cite{SRC}. Unless one puts by hand the 
spin-rotation coupling term into the Hamiltonian of the system 
or takes into account the higher order spin relativistic effects 
in a non-relativistic approximation \cite{SRC}, this is not the 
case because the Hawking radiation radiation \cite{Hawk} spectra 
show that there is no such thing in a stationary rotating
black hole background. The well-known thermal radiation spectra  
\cite{DRS,WC1} of all particle species in a stationary space-time 
are given by 
\be 
\langle {\cal N}_{\omega} \rangle \sim \{\exp[(\omega 
-m\Omega_h -q\Phi_h)/T_h] \pm 1\}^{-1} \, , \label{trs}
\ee
where three constants $\Omega_h, \Phi_h, T_h$ are, respectively, 
the angular velocity, the electromagnetic potential and the 
effective temperature of the event horizon of the hole, while 
$m$ is the azimuthal quantum number of the particle, $q$ its 
charge. The spectrum formula of Eq. (\ref{trs}) can be easily 
derived from the Teukolsky master equation \cite{Teuk} in the 
Kerr space-time \cite{Kerr} by means of the method suggested by 
Damour-Ruffini and Sannan \cite{DRS}.

Situation may be changed if one turns to consider the Hawking 
radiation of particles with higher spin in a dynamic black hole.
In recent papers \cite{WC2}, the evaporation of Dirac particles 
in a variable-mass Kerr black hole \cite{GHJW,CKC} has been 
discussed by use of a method of generalized tortise coordinate 
transformation (GTCT). A new interaction effect due to the 
coupling of the intrinsic spin of Dirac particles with the 
angular momentum of the radiating Kerr black hole was observed 
in the thermal radiation spectrum of particles with spin-$1/2$. 
The character of this spin-rotation coupling effect is its obvious 
dependence on different helicity states of the spinning particles. 
This effect disappears \cite{WC3} when the space-time degenerates 
to a spherically symmetric black hole of Vaidya-type. It should be 
noted that this effect displayed in the Fermi-Dirac spectrum is 
absent in the Bose-Einstein distribution of Klein-Gordon particles.

The aim of this letter is twofold. First, we deduce the perturbed 
wave equations for scalar, neutrino and electromagnetic fields in 
an evaporating Kerr black hole \cite{GHJW,CKC}. It should be noted 
that the Teukolsky's context \cite{Teuk} on the perturbations of 
the Kerr black hole \cite{Kerr} can not apply here to the gravitational 
perturbation of this space-time because it is of Petrov type-II. 
We point out that the black hole radiation of the Vaidya metric
has already been studied \cite{KCKY} within the Teukolsky's celebrated 
framework \cite{Teuk}. Second, we deal with the thermal radiation of 
particles with spin-$0, 1/2$ and $1$ in the non-stationary Kerr black 
hole. The method used here is just the same as that we have developed 
in \cite{WC2} to investigate the Hawking effect of Dirac particles in 
this space-time. We find that in the Hawking radiation spectra of 
spinning particles, there appears an extra interaction energy due 
to the coupling of the intrinsic spin of particles with the rotation 
of the black hole. This effect vanishes when the space-time becomes 
a stationary Kerr black hole or a Vaidya-type spherically symmetric 
black hole.

The metric describing a radiating Kerr black hole can be written 
in the advanced Eddington-Finkelstein coordinates system as \cite{GHJW,CKC}
\bea
ds^2  
&=& \frac{\Delta -a^2\sda}{\Sigma}dv^2 
-2dvdr +2a\sda drd\varphi  \nn\\
&-& \Sigma d\theta^2 +2\frac{r^2 +a^2 
-\Delta}{\Sigma}a\sda dvd\varphi \nn\\
&-& \frac{\big(r^2 +a^2\big)^2 
-\Delta a^2\sda}{\Sigma}\sda d\varphi^2 \, , 
\eea
where $\Delta = r^2 -2M(v)r +a^2$, $\Sigma = r^2 +a^2\cda = \rho^*\rho$, 
$\rho = r +ia\cta$, and $v$ is the standard advanced time. The line 
element is a natural non-stationary generalization of the stationary 
Kerr solution. The mass $M$ depends on the time $v$, but the specific 
angular momentum $a$ is a constant.

The space-time geometry of an evaporating Kerr black hole is 
characterized by three kinds of surfaces of particular interest: 
the apparent horizons $r_{AH}^{\pm} = M\pm\sqrt{M^2 -a^2}$, the 
timelike limit surfaces $r_{TLS}^{\pm} = M\pm\sqrt{M^2-a^2\cta}$, 
and the event horizons $r_{EH}^{\pm} = r_H$. The event horizon 
is necessary a null-surface $r = r(v,\theta)$ that satisfies the 
null-surface conditions $g^{\mu\nu}\p_{\mu} F\p_{\nu} F = 0$ and 
$F(v,r,\theta) = 0$. 

The traditional method to determine the location and the temperature 
of the event horizon of a dynamic black hole is to calculate vacuum 
expectation value of the renormalized energy momentum tensor \cite{YB}. 
But this method is very complicated and meets great difficulties here. 
A more effective method is called the generalized tortoise coordinate 
transformation (GTCT) which can give simultaneously the exact values 
both of the location and of the temperature of the event horizon of 
a non-stationary black hole. Basically, this method is to reduce 
Klein-Gordon or Dirac equation in a known black hole background to a 
standard wave equation near the event horizon by generalizing the common 
tortoise-type coordinate $r_* = r +\frac{1}{2\kappa}\ln(r -r_H)$ 
in a static or stationary space-time \cite{DRS} (where $\kappa$ is the 
surface gravity of the studied event horizon) to a similar form in a 
non-static or non-stationary space-time and by allowing the location 
of the event horizon $r_H$ to be a function of the advanced time 
$v = t +r_*$ and/or the angles $\theta,\varphi$. 

As the space-time under consideration is symmetric about 
$\varphi$-axis, one can introduce the following generalized 
tortoise coordinate transformation (GTCT) \cite{WC2} 
\bea
&&r_* = r +\frac{1}{2\kappa}\ln\big[r -r_H(v,\theta)\big] \, , \nn\\
&&v_* = v -v_0 \, , ~~~\theta_* = \theta -\theta_0 \, , 
\label{trans}
\eea
where $r_H = r(v,\theta)$ is the location of event horizon, and $\kappa$ 
is an adjustable parameter. All parameters $\kappa$, $v_0$ and $\theta_0$ 
characterize the initial state of the hole and are constant under the 
tortoise transformation.

Substituting the GTCT of Eq. (\ref{trans}) into the null-surface equation 
$g^{\mu\nu}\p_{\mu} F\p_{\nu} F = 0$ and then taking the $r \rightarrow 
r_H(v_0, \theta_0)$, $v \rightarrow v_0$ and $\theta \rightarrow 
\theta_0$ limits, we can arrive at
\be
\Big[\Delta_H -2\big(r_H^2 +a^2\big)\dot{r}_H +a^2\sda_0 \dot{r}_H^2 
+\pdrH\Big]\Big(\frac{\p F}{\p r_*}\Big)^2 = 0 \, ,
\ee
in which the vanishing of the coefficient in the square bracket can
give the following equation to determine the location of the event 
horizon of an evaporating Kerr black hole
\be
\Delta_H -2\big(r_H^2 +a^2\big)\dot{r}_H 
+a^2\sda_0 \dot{r}_H^2 +\pdrH  = 0 \, , 
\label{loeh}
\ee
where we denote $\Delta_H = r_H^2 -2M(v_0)r_H +a^2$. The quantities 
$\dot{r}_H = \pv r_H$ and $\prH = \pta r_H$ depict the change of the 
event horizon in the advanced time and with the angle, which reflect 
the presence of quantum ergosphere near the event horizon. 
Eq. (\ref{loeh}) means that the location of the event horizon 
is shown as
\bea
r_H &=& \big\{M \pm \big[M^2 -\big(a^2\sda_0\dot{r}_H^2 
+\pdrH\big)\big(1 -2\dot{r}_H\big) \nn\\
&-& a^2\big(1 -2\dot{r}_H\big)^2\big]^{1/2}\big\}/(1 -2\dot{r}_H) \, . 
\label{loca}
\eea

To derive the scalar, neutrino and electromagnetic 
perturbations in the evaporating Kerr black hole, 
we establish a complex null-tetrad which has 
$[v, r, \theta, \varphi]$ components:
$l^{\mu} = -\delta_1^{\mu}$,  
$n^{\mu} = \big[\big(r^2 +a^2\big)\delta_0^{\mu} 
+2^{-1}\Delta\delta_1^{\mu} +a\delta_3^{\mu}\big]/\Sigma$, 
$m^{\mu} = \big(ia\sta \delta_0^{\mu} +\delta_2^{\mu} 
+i\delta_3^{\mu}/\sta\big)/\big(\sqd\rho\big)$.
Within the Newman-Penrose (NP) formalism \cite{NP}, 
we substitute, respectively, 
$\chi_1 = \sqrt{2\Sigma}\eta_1 \, , 
\chi_0 = \sqrt{\Sigma}\eta_0/\rho^* $ 
for the Weyl spinors $\eta_0, \eta_1$ and 
$\Phi_0 = \rho\phi_0/\big(\sqd\rho^*\big) \, ,
\Phi_1 = \rho\phi_1 \, , \Phi_2 = \sqd\Sigma\phi_2$ 
for the Maxwell complex scalars $\phi_0, \phi_1, \phi_2$
into the NP forms of both Weyl equation and Maxwell equation 
\cite{Teuk}, and get the following first-order equations
\be
\pr\chi_1 +{\cL}_{\hf}\chi_0 = 0 \, , 
~~\Delta {\cD}_{\hf}\chi_0 -{\cLd}_{\hf}\chi_1 = 0 \, , 
\label{foen}
\ee
for the Weyl neutrinos, and
\bea
&&\big(\pr +1/\rho\big) \Phi_1 
+\big({\cL}_1 +ia\sta/\rho\big)\Phi_0 = 0 \, , \nn \\
&&\Delta \big({\cD}_1 -1/\rho\big)\Phi_0 
-\big({\cLd}_0 -ia\sta/\rho\big)\Phi_1 = 0 \, , \nn \\
&&\big(\pr -1/\rho\big) \Phi_2 
+\big({\cL}_0 -ia\sta/\rho\big)\Phi_1 = 0 \, , \nn\\
&&\Delta \big({\cD}_0 +1/\rho\big)\Phi_1 
-\big({\cLd}_1 +ia\sta/\rho\big)\Phi_2 \nn\\
&&~~= 2i\dot{M}ra\sta\Phi_0 \, ,
\label{foep}
\eea
for the photons. Here we have defined operators
${\cD}_n = \pr +2\Delta^{-1}\left[n(r -M) +a\pvi 
+\big(r^2 +a^2\big)\pv\right]$,
${\cL}_n = \pta +n\coa -\frac{i}{\sta}\pvi -ia\sta\pv$,
and the complex conjugate ${\cLd}_n$ for operator ${\cL}_n$.  

Eqs. (\ref{foen},\ref{foep}) can not be decoupled 
except in the stationary Kerr black hole case 
($M= const$) or in a Vaidya-type space-time. However, 
to deal with the problem of Hawking radiation, one 
may concern about their asymptotic behaviors near 
the horizon only. Making use of the transformation of 
Eq. (\ref{trans}) and then taking the $r \rightarrow 
r_H(v_0,\theta_0)$, $v \rightarrow v_0$ and $\theta 
\rightarrow \theta_0$ limits, one can reduce Eqs. 
(\ref{foen},\ref{foep}) to the following forms 
\bea
&&\frac{\p\Psi_{p+1}}{\p r_*} 
-\Big(\prH -ia \sta_0\dot{r}_H \Big)\frac{\p\Psi_p}{\p r_*} = 0 \, ,
\nn\\ 
&&\Big(\prH +ia\sta_0\dot{r}_H \Big)\frac{\p\Psi_{p+1}}{\p r_*} \nn\\
&&~~ +\Big[\Delta_H -2\big(r_H^2+a^2\big) \dot{r}_H \Big] 
\frac{\p\Psi_p}{\p r_*} = 0 \, ,  
\label{wone}
\eea
where $\Psi_p$ stands for the Weyl spinors $\chi_0, \chi_1$ 
when $p = 0, 1 (s = 1/2)$ and represents the Maxwell scalars 
$\Phi_0, \Phi_1, \Phi_2$ when $p = 0, 1, 2 (s = 1)$. The 
existence condition of nontrial solutions for $\Psi_p$ is 
that the determinant of Eq. (\ref{wone}) vanishes which 
results in the event horizon equation (\ref{loeh}) given above.  

With the first-order equations (\ref{foen}, \ref{foep}) in 
hand, we are now in a position to derive their corresponding 
second-order equations
\bea
&&\big(\pr\Delta {\cD}_{\hf} +{\cLd}_{1/2}{\cL}_{\hf}\big)\chi_0 = 0 \, ,
\nn\\ 
&&\big(\Delta {\cD}_{\hf}\pr +{\cL}_{\hf}{\cLd}_{\hf}\big)\chi_1 \nn\\ 
&&~~ = ia\sta \big(2\dot{M}r\pr +\dot{M} \big)\chi_0  \, , \label{soen}
\eea
and
\bea
&&\big(\pr\Delta {\cD}_1 +{\cLd}_0{\cL}_1 +2\rho^*\pv\big)\Phi_0 = 0 \, , 
\nn\\
&&\big(\pr\Delta {\cD}_0 +{\cLd}_1{\cL}_0 -2\rho^*\pv
+2M\rho^*/\rho^2\big)\Phi_1 \nn\\
&&~~ = 2i\dot{M}ra\sta \big(\pr -1/\rho\big)\Phi_0 \, , \nn\\
&&\big(\Delta {\cD}_0\pr +{\cL}_0{\cLd}_1 -2\rho^*\pv\big)\Phi_2 \nn\\
&&~~ = -2\ddot{M}ra^2\sda\Phi_0 -4i\dot{M}ra\sta{\cL}_1\Phi_0 \, . 
\label{soep}
\eea
The source-less wave equation $\Box\Phi = 0$ for 
a massless scalar field can be written as 
\be
\big(\Delta {\cD}_1\pr +{\cL}_1{\cLd}_0 
+2\rho^*\pv\big)\Phi = 0 \, . \label{soes}
\ee
These equations can be thought of as the generalized Teukolsky master
equations, they encompass the well-known results when the mass of the 
black hole is a constant \cite{Teuk} or the black hole is non-rotating 
\cite{KCKY}. Because the radiating Kerr metric is of Petrov type-II, 
the gravitational perturbations of this space-time is more involved.
It can be done for the $\psi_0, \psi_1$ components but not easily 
for the other components of the Weyl tensors. 

Having derived the master equations controlling the scalar, neutrino 
and electromagnetic perturbations of the non-stationary Kerr metric, we 
are now ready to study the quantum thermal property of this space-time
by investigating the Hawking radiation of massless particles with 
spin-$0, 1/2$ and $1$. Given the GTCT in Eq. (\ref{trans}), the limiting 
form of Eqs. (\ref{soen}-\ref{soes}), when $r$ approaches $r_H(v_0, 
\theta_0)$, $v$ goes to $v_0$ and $\theta$ goes to $\theta_0$, can be 
recast into the standard wave equation near the event horizon in an 
united form 
\bea
&&\frac{\p^2\Psi_p}{\p r_*^2} +2\frac{\p^2\Psi_p}{\p r_* \p v_*}
+2\Omega\frac{\p^2\Psi_p}{\p r_* \p\varphi} \nn\\
&&~ +2C_3\frac{\p^2\Psi_p}{\p r_*\p\theta_*} 
+2\big(C_2 +iC_1\big)\frac{\p\Psi_p}{\p r_*} = 0 \, , 
\label{wtwo}
\eea
where 
$\Omega$, $C_3$, $C_2$ and $C_1$ are all real constants,
\bea
&&\Omega = a(1 -\dot{r}_H)/\big[r_H^2 
+a^2 -\dot{r}_H a^2\sda_0\big] \, , \nn\\
&&C_3 = -\prH/\big[r_H^2 +a^2 -\dot{r}_H a^2\sda_0\big] \, , \nn\\
&&C_2 +iC_1 = \frac{-1}{2\big(r_H^2 +a^2 
-\dot{r}_H a^2\sda_0\big)}\Big[2p\big(r_H -M \nn\\
&& +ia\cta_0\dot{r}_H\big) -2(2p +1)r_H\dot{r}_H 
+\ddot{r}_H a^2\sda_0 +\pprH \nn\\ 
&& +\coa_0\prH +2(s +p)\dot{M}r_H\frac{a^2\sda_0\dot{r}_H 
-ia\sta_0\prH}{\Delta_H 
-2\dot{r}_H \big(r_H^2+a^2\big)}\Big] \, .\nn
\eea
The wave equation (\ref{wtwo}) includes the scalar field $\Phi$ 
as a special case when $p = s =0$. Also it holds true for a 
propagating external gravitational field with suitable replacements 
of all field components. 

In deriving Eq. (\ref{wtwo}), we have dealt with an infinite 
form of $0/0$-type in order to obtain a finite value 
$2(r_H -M) -4r_H\dot{r}_H$ and adjusted the parameter $\kappa$ 
such that it satisfies
\bea
&&\frac{r_H\big(1 -2\dot{r}_H\big) -M}{\kappa}
+2\Delta_H -2\dot{r}_H\big(r_H^2+a^2\big) \nn\\
&&~~ = r_H^2 +a^2 -\dot{r}_H a^2\sda_0 \, ,
\eea
which means the surface gravity is
\bea
\kappa = \frac{r_H\big(1-2\dot{r}_H\big) -M}{\big(r_H^2 
+a^2 -\dot{r}_H a^2\sda_0\big)\big(1 -2\dot{r}_H\big) 
+2\pdrH} \, . \label{temp}
\eea
The location (\ref{loca}) and the temperature $\kappa/(2\pi)$
of event horizon of a non-stationary Kerr black hole are 
shown to be dependent not only on the advanced time $v$ but 
also on the angle $\theta$. Both of them can be calculated 
by iteration, but we don't pursue this goal here. A crucial step 
of our treatment is to use the relations between the first-order 
derivatives (\ref{wone}) to eliminate the crossing-term of the 
first-order derivatives in the second-order equations near the 
event horizon. It is also important to adjust the parameter 
$\kappa$ so as to recast each second-order equation into a 
standard wave equation near the event horizon. 

Now that all real coefficients in Eq. (\ref{wtwo}) can be regarded as 
finite constants, one can separate variables as $\Psi_p = R(r_*)
\exp[\lambda \theta_* +i(m\varphi -\omega v_*)]$ and have a solution 
to the radial part as $R = R_1\exp[2i(\omega -m\Omega -C_1 +iC_0)r_*] 
+R_0$, in which $\lambda$ is a real separation constant, $C_0 = 
\lambda C_3 +C_2$. The ingoing wave $\Psi_p^{\rm in} \sim \exp[\lambda 
\theta_* +i(m\varphi -\omega v_*)]$ is regular at the event horizon 
$r = r_H$, whereas the outgoing wave 
\be
\Psi_p^{\rm out}(r > r_H) = \Psi_p^{\rm in}\exp[2i(\omega 
-m\Omega -C_1 +iC_0)r_*] \, , 
\ee
is irregular, it can be analytically continued from the outside of 
the hole into the inside of the hole through the lower complex 
$r$-plane to 
\be
\widetilde{\Psi_p^{\rm out}}(r < r_H) 
= \Psi_p^{\rm out}\exp[\pi(\omega -m\Omega -C_1 +iC_0)/\kappa] \, .
\ee
According to the method of Damour-Ruffini-Sannan's \cite{DRS}, 
the relative scattering probability at the event horizon and the 
thermal radiation spectrum of particles from the black hole are,
respectively,
\bea
&&\Big|\frac{\Psi_p^{\rm out}}{\widetilde{\Psi_p^{\rm out}}}\Big|^2
= \exp[-2\pi(\omega -m\Omega -C_1)/\kappa] \, , \nn\\
&&\langle {\cal N}(\omega) \rangle = 
\frac{\Gamma(\omega)}{\exp[2\pi(\omega -m\Omega 
-C_1)/\kappa] -(-1)^{4s^2}} \, , \label{sptr}
\eea
where $\Gamma(\omega)$ is the transmission coefficient in 
certain modes with which a particle can escape from the event 
horizon to infinity, $m$ the azimuthal quantum number. $\Omega$ 
can be interpreted as the angular velocity of the event horizon 
of the evaporating Kerr black hole, while the explicit expression 
of the \lq\lq spin-dependent" term $C_1$ reads 
\bea
C_1 &=& \frac{1}{r_H^2 +a^2 -\dot{r}_H a^2\sda_0}
\Big[-pa\cta_0\dot{r}_H \nn\\
&+&~ (s +p)\frac{\dot{M}r_H a\sta_0\prH}{\Delta_H 
-2\dot{r}_H\big(r_H^2 +a^2\big)}\Big] \, .
\eea

The thermal radiation spectrum (\ref{sptr}) of particles with 
spin-$s$ is composed of two parts: $\omega_p = m\Omega +C_1$, 
one is the rotational energy $m\Omega$ arising from the coupling 
of the orbital angular momentum of particles with the rotation 
of the black hole; another is $C_1$ due to the coupling of the 
intrinsic spin of particles and the angular momentum of the hole, 
which vanishes in the case of a stationary Kerr black hole ($M = 
const$, $\dot{r}_H = \prH = 0$) or a Vaidya-type black hole ($a 
= 0$, $\prH = 0 $, $\dot{r}_H \not= 0$). The distribution (\ref{sptr}) 
recovers the known results \cite{WC3,KCKY} when the black hole is 
non-rotating, and approximates the formula (\ref{trs}) when $\dot{r}_H 
\simeq 0$ (namely, the luminosity $L = -dM/dv = -\dot{M} \simeq 0$). 
To see its meaning more transparently, let us, of astrophysical 
interest, neglect the angular deformation $\prH$ of the hole, 
then we can approximate $\omega_p$ as
\be
\omega_p \simeq \big(m -p \frac{\dot{r}_H}{1 -\dot{r}_H} 
\cta_0\big)\Omega \, , ~~ p = 0, 1, \cdots, 2s
\ee
in the case of small evaporation and slow rotation. The factor 
$\dot{r}_H/(1- \dot{r}_H)$ describes the evolution of the 
hole in the time, while the factor $\cta_0$ comes from the scalar 
product of the spin-rotation coupling. The term $C_1$ is obviously 
related to the helicity of particles in different spin states, it 
characterizes a new effect arising from the interaction between 
the spin of particles and the rotation of a evaporating black hole. 
The feature of this new effect is that it is dependent on different 
helicity states of spinning particles. 

In summary, this study not only encompasses the thermal spectrum 
of spinning particles in the Vaidya metric \cite{KCKY}, but also
provides a partial confirmation of York's conjecture \cite{YB} that
quantum radiance of a charged and rotating hole might be 
originated from the quantum ergosphere effect. Here we suggest that
the radiative mechanism of an evaporating Kerr black hole can be
changed by the quantum rotating ergosphere which can be viewed as 
a mixture of the classical rotating ergosphere and York's quantum
ergosphere. We argue that dynamic black holes must have some 
new properties very different from that of stationary ones, and 
the spin-rotation coupling effect presented in this study may be 
a good example. We suggest a new experiment in which this effect 
can be observed from the spectrography of certain astrophysical 
objects in the future.

This work is supported in part by the NSFC in China.

\noindent
$^*$E-mail: sqwu@iopp.ccnu.edu.cn \\ 
$^{\dagger}$E-mail: xcai@ccnu.edu.cn


\begin{references}

\bibitem{CR}
D. Christodoulou, Phys. Rev. Lett. {\bf 25} (1970) 1596;
D. Christodoulou and R. Ruffini, Phys. Rev. {\bf D4} (1971) 3552.  

\bibitem{Smar}
L. Smarr, Phys. Rev. Lett. {\bf 30} (1973) 71, 521(E).

\bibitem{SRC}
Y. N. Obukhov, Phys. Rev. Lett. {\bf 86} (2001) 192;
B. Mashhoon, Class. Quant. Grav. {\bf 17} (2000) 2399; 
Gel. Rel. Grav. {\bf 31} (1999) 681;
B. Mashhoon, R. Neutze, M. Hannam and G. E. Stedman, Phys. Lett.
{\bf A249} (1998) 161;
L. H. Ryder and B. Mashhoon, {\it Spin and Rotation in General
Relativity}, gr-qc/0102101.

\bibitem{Hawk}
S. W. Hawking, Nature, {\bf 248} (1974) 30; Commun. Math. Phys. {\bf 43}
(1975) 199.

\bibitem{DRS} 
T. Damour and R. Ruffini, Phys. Rev. {\bf D14} (1976) 332;
S. Sannan, Gen. Rel. Grav. {\bf 20} (1988) 239.

\bibitem{WC1}
S. Q. Wu and X. Cai, IL Nuovo Cimento, {\bf B115} (2000) 143;
Int. J. Theor. Phys. {\bf 39} (2000) 2215.

\bibitem{Teuk}
S. A. Teukolsky, Phys. Rev. Lett. {\bf 29} (1972) 1114;
S. A. Teukolsky, Astrophys. J. {\bf 185} (1973) 635; 
W. Press and S. A. Teukolsky, {\it ibid}. {\bf 185} (1973) 649;
S. A. Teukolsky and W. Press , {\it ibid}. {\bf 193} (1974) 443.
 
\bibitem{Kerr}
R. P. Kerr, Phys. Rev. Lett. {\bf 11} (1963) 237.

\bibitem{WC2}
S. Q. Wu and X. Cai, Chin. Phys. Lett. {\bf 18} (2001) 485; {\it Hawking 
radiation of Dirac particles in a variable-mass Kerr space-time}, to appear 
in Gen. Rel. Grav. {\bf 33}, No. 7 (2001).

\bibitem{GHJW}
C. Gonzalez, L. Herrera, and J. Jimenez, J. Math. Phys. {\bf 20} (1979) 837;
J. L. Jing and Y. J. Wang, Int. J. Theor. Phys. {\bf 35} (1996) 1481. 

\bibitem{CKC}
M. Carmeli and M. Kaye, Ann. Phys. (NY) {\bf 103} (1977) 197;
M. Carmeli, {\sl Classical Fields: General Relativity and Gauge Theory}, 
(New York: John Wiley \& Sons, 1982).  

\bibitem{WC3}
S. Q. Wu and X. Cai, {\it Asymmetry of Hawking Radiation of Dirac Particles 
in a Charged Vaidya - de Sitter Black Hole}, to apear in Int. J. Theor. Phys. 
{\bf 40}, No. 7 (2001).

\bibitem{KCKY}
S. W. Kim, E. Y. CHOI, S. K. Kim and J. Yang, Phys. Lett. {\bf A141} (1989)
238.

\bibitem{YB}
J. W. York, Jr., Phys. Rev. {\bf D28} (1983) 2929;
R. Balbinot, {\it ibid}. {\bf D33} (1986) 1611.
 
\bibitem{NP} 
E. Newman and R. Penrose, J. Math. Phys. {\bf 3} (1962) 566; 
S. Chandrasekhar, {\sl The Mathematical Theory of Black Holes}, 
(New York: Oxford University Press, 1983).

\end{references}
\end{document}